\documentclass[%
 aps,
superscriptaddress,
groupedaddress,
showpacs,
 amsmath,amssymb,
pre,
twocolumn
]{revtex4-1}

\usepackage{graphicx}  % needed for figures
\usepackage{dcolumn}   % needed for some tables
\usepackage{bm}        % for math

\begin{document}

%\preprint{APS/123-QED}

\title{Iterative control strategies for non--linear systems.\\}% Force line breaks with \\

\author{G.~Forte}
\email{gforte@nd.edu}
 \affiliation{University of Notre Dame, Department of Physics, 225 Nieuwland Science Hall, Notre Dame, IN 46556 USA}%Lines break automatically or 
 %can be forced with \\
\author{D.~C.~Vural}%
\email{dvural@nd.edu}
\affiliation{University of Notre Dame, Department of Physics, 225 Nieuwland Science Hall, Notre Dame, IN 46556 USA}%


%\affiliation{
%\affiliation{
 %Third institution, the second for Charlie Author
%\affiliation{%


\date{\today}% It is always \today, today,

\begin{abstract}
In this paper, we focus on the control of the mean field equilibrium of non linear networks of the Langevin type in the limit of small noise. Using iterative linear approximations, we derive a formula that prescribes a control strategy in order to displace the equilibrium state of a given system, and remarkably find that the control function has a ``universal'' form under certain physical conditions. This result can be employed to define universal protocols useful, for example, in the optimal work extraction from a given reservoir.  Generalizations and limits of application of the method are discussed.

%\begin{description}
%\end{description}
\end{abstract}

\pacs{Valid PACS appear here}% PACS, 
\maketitle

%*************************
\section{Introduction.}
\label{sec:intro}
%*************************
Thermodynamics of finite--size systems is a rapidly expanding field. From the theoretical side, a number of non--equilibrium relations have been established \cite{van2015,seifert2008,jarzynski1997,hatano2001,marconi2008}, starting with the pioneering work of \citet{evans1993}. In the mean time, experimental techniques were developed to probe mesoscopic systems such as colloidal particles and macro--molecules, allowing for the experimental validation of some of these non--equilibrium thermodynamic relations \cite{blickle2006,trepagnier2004,hayashi2010,liphardt2002,marconi2008}. 
Present control methods are based on either finding the control function $\lambda(t)$ \cite{bellman1964,astrom2012,wang2002,aurell2011} or the optimal probability density function governing $\lambda(t)$ \cite{ljung1998, abreu2011} in order to attain a certain mean or variance. Controlling rare events is another problem of relevance, addressed either by path integration methods \cite{kappen2005,wells2015,kappen2005_1} or the WKB approximation \cite{assaf2010,khasin2011}. An interesting application of the WKB method to reaction--diffusion systems has been developed in \cite{elgart2004,elgart2006}. 

Here we focus on the problem of controlling non--linear systems with finite degrees of freedom subject to weak noise. Our control scheme involves driving a system from one equilibrium state to another by an external protocol, while minimizing some cost functional. A first step towards determining an optimal protocol in stochastic thermodynamics has been taken by \citet{schmiedl2007}, who considered linear systems, and minimized the physical work performed by the control force on the system.  Here we generalize this result by obtaining expressions prescribing the control of the mean field equilibrium of \emph{non--linear} networks, constrained by a \emph{generic} cost functional. More interestingly, we find a special limiting behavior for which the optimal protocol becomes independent of the particular form on the cost functional, i.e. taking a universal form, and thus, allowing for the design of universal optimal controllers.

A general, multi--dimensional stochastic system can be written as
\begin{equation}
\frac{d \mathbf{q}}{dt} = \mathbf{F}(\mathbf{q}) + \sqrt{2}\boldsymbol{\xi}(t) +\boldsymbol{\lambda}(t)
\label{eq:gensys}
\end{equation}
where $\boldsymbol{\lambda}(t)$ is a control parameter and $\boldsymbol{\xi}$ is the standard white noise, i.~e. $\langle \boldsymbol{\xi}(t)\rangle= 0$, $\langle \xi_{i}(t)\xi_{j}(t')\rangle = \delta_{ij}\delta(t-t')$. We take the noise intensity to be unity, which is always possible after rescaling $t$ and $\mathbf{F}(\mathbf{q})$. Eq.~\eqref{eq:gensys} is general and can be used to describe a broad class of physical, biological and/or ecological systems. 

Our goal is to transition a system initially occupying an equilibrium state $\mathbf{q}^{*}_{0}$ to $\mathbf{q}^{*}_{0} + \mathbf{C}$, using a control function that varies from $\boldsymbol{\lambda}(t=0)=\boldsymbol{\lambda}_{s}$ 
to $\boldsymbol{\lambda}(\tau) = \boldsymbol{\varepsilon}$ within some finite time $\tau$. Furthermore, we would like the control to be \emph{optimal}. This means that our protocol $\boldsymbol{\lambda}(t)$ must minimize some cost functional,
\begin{equation}
\mathcal{W} = \int_{0}^{\tau} dt\mbox{ }\mathcal{L}(\boldsymbol{\lambda},\dot{\boldsymbol{\lambda}},\mathbf{q},\dot{\mathbf{q}}).
\label{eq:CostOne}
\end{equation}

Here, we consider a general class of Lagrangians of the form
\begin{equation}
\mathcal{L}=\langle	\Gamma(\mathbf{q},\dot{\mathbf{q}},\boldsymbol{\lambda},\dot{\boldsymbol{\lambda}})\rangle
\label{eq:lagrangian}
\end{equation}

We do not assume an explicit form for $\mathbf{F}(\mathbf{q})$ and $\Gamma(\mathbf{q},\dot{\mathbf{q}},\boldsymbol{\lambda},\dot{\boldsymbol{\lambda}})$. However, we assume that $M_{ij}(\mathbf{\boldsymbol{\lambda}})$ is a simply connected differentiable manifold, whose real parts of eigenvalues remain negative for the range of $\boldsymbol{\lambda}$ to be applied. This ensures that the shift in equilibrium $\mathbf{q}^{*}(\boldsymbol{\lambda})$ takes place smoothly, as $\boldsymbol{\lambda}$ is varied.

The paper is organized as follow: In Sec.~\ref{sec:main} we introduce the main idea. In Sec.~\ref{sec:spec} we evaluate the protocol for special limits. In particular, in Sec.\ref{sec:univ} we identify a special class of Lagrangians such that the control protocol is universal, i.e. \emph{independent} of the Lagrangian parameters. In Sec.~\ref{sec:gen} we generalize the formulas derived in Sec.~\ref{sec:main}. In Sec.~\ref{sec:disc} we consider some explicit examples and discuss our limitations. Finally we present concluding remarks in Sec.~\ref{sec:concl}.

%********************************
\section{Iterative solution to the non linear control problem.}
\label{sec:main}
%********************************
We start our analysis with small displacements, $|(\mathbf{q}^{*}_{0} + \mathbf{C})|/|\mathbf{q}_{0}^{*}|\sim 1$. That is, we move in the neighborhood of the original equilibrium. This condition will be relaxed later by iterating over many such small displacements. 

To simplify our notation we will shift our origin to the original equilibrium state, i.~e. $\mathbf{q}\to \mathbf{q}-\mathbf{q}^{*}_{0}$. Near this point, the system can be linearized, $\dot{q}_{i} \approx -\sum_{j}M_{ij}  q_{j}  + \lambda_{j}(t) +\sqrt{2}\xi_{j}(t)$, where $M_{ij}=M_{ij}(\mathbf{0})$ is (minus) the stability matrix evaluated around the original equilibrium. Thus, for small displacements, the problem is reduced to controlling a linear system, starting from $\boldsymbol{\lambda}(t=0) = \mathbf{0}$ 
and ending at $\boldsymbol{\lambda}(\tau) = \boldsymbol{\varepsilon}$ in the finite time $\tau$.   
Since we aim to perform a small displacement, we can also expand the Lagrangian $\mathcal{L}$ around $\boldsymbol{\lambda},\dot{\boldsymbol{\lambda}}\sim \mathbf{0}$. Thus, omitting constant terms, the expanded Lagrangian has the form
\begin{equation} 
\mathcal{L} = 
\langle \mathbf{h}(\mathbf{q,\dot{\mathbf{q}}})\cdot\boldsymbol{\lambda} \rangle + \langle \mathbf{g}(\mathbf{q,\dot{\mathbf{q}}})\cdot\dot{\boldsymbol{\lambda}} \rangle
\label{eq:lagra2}
\end{equation}
The expansion around $\boldsymbol{\lambda},\dot{\boldsymbol{\lambda}}\sim \mathbf{0}$ automatically implies an expansion around the equilibrium and thus we can take
\begin{align}
\mathbf{g}&\approx \sum_{i=1}^{N}(\gamma^{i}Q_{i} + \eta^{i}\dot{Q}_{i}) + \mbox{const.}
\label{eq:g}\\
\mathbf{h}&\approx \sum_{i=1}^{N}(\nu^{i}Q_{i} + \alpha^{i}\dot{Q}_{i}) + \mbox{const.}
\label{eq:h}\\
\mathcal{L} &\approx \sum_{i=1}^{N}(\nu^{i} Q_{i}\lambda_{i} + \alpha^{i} \dot{Q}_{i}\lambda_{i} + \gamma^{i} Q_{i}\dot{\lambda}_{i} + \eta^{i} \dot{Q}_{i}\dot{\lambda}_{i})\label{eq:lagrapp}
\end{align}
where $Q_{i} = \langle q_{i}\rangle$ is the mean trajectory and $\nu^{i}, \alpha^{i}, \gamma^{i}, \eta^{i}$ are expansion coefficients. 
Eq.~\eqref{eq:lagra2} is the most general form for a Lagrangian up to the first order in $\boldsymbol{\lambda}, \dot{\boldsymbol{\lambda}}$ and can be used to describe any system near its equilibrium. In Sec.~\ref{sec:univex} we discuss some specific examples and point out to physically--relevant Lagrangians that fall under this form. 

Plugging Eq.~\eqref{eq:lagra2} in Eq.~\eqref{eq:CostOne}, obtaining $\lambda_{i}$ and $\dot{\lambda}_{i}$ from the mean--field equation 
$\dot{Q}_{i} = -\sum_{j}M_{ij} Q_{j}  + \lambda_{i}(t)$ and omitting the terms that do not enter the minimization explicitly, we obtain (for a full derivation cf.~Appendix~\ref{sec:one}) 

\begin{align}
\mathcal{W} = \sum_{i=1}^{N}\left\{\gamma^{i}\Bigl[Q_{i}\dot{Q}_{i}\Bigr]_{0}^{\tau}+\frac{\eta^{i}}{2}\Bigl[\dot{Q}_{i}^{2}\Bigr]_{0}^{\tau} +
\right.\nonumber
\\
+M_{ii}\int_{0}^{\tau}dt\mbox{ }
(a^{i}\dot{Q}_{i}^{2} + \nu^{i}Q_{i}^{2}) +\nonumber
\\
\left.\sum_{j\ne i}M_{ij}\int_{0}^{\tau}dt\Bigl(\nu^{i}Q_{i}Q_{j} +\eta^{i}\dot{Q}_{i}\dot{Q}_{j}\Bigr)\right\}
\label{eq:WFIN}
\end{align}
where 
\begin{equation}
a^{i}  =   \eta^{i} + \frac{\alpha^{i}- \gamma^{i}}{M_{ii}}.
\label{eq:ai}
\end{equation}
Using Euler--Lagrange equations to minimize this functional, we obtain
\begin{equation}
\sum_{j=1}^{N}(A_{ij}\ddot{Q}_{j}+B_{ij}Q_{j})=0,
\label{eq:ODEs}
\end{equation}
where the matrices $A_{ij}$ and $B_{ij}$ are
\begin{equation}
A_{ij}=2a^{j}\delta_{ij}-\frac{M_{ij}}{M_{jj}}\eta^{i}+\frac{M_{ij}}{M_{jj}}\delta_{ij}\eta^{i}
\label{eq:A0}
\end{equation}
\begin{equation}
B_{ij}=-2\nu^{j}\delta_{ij}-\frac{M_{ij}}{M_{jj}}\nu^{i}+\frac{M_{ij}}{M_{jj}}\delta_{ij}\nu^{i}
\label{eq:B0}
\end{equation}
Since the system Eq.~\eqref{eq:ODEs} is linear and homogeneous, the general solution is of the form $Q^{\beta}_{j}=\sum_{k}\beta_{jk}\phi_{k}(t)$, where $\beta_{jk}$ are constant coefficients and the $\phi_{k}(t)$ are of the form given in Eq.~\eqref{eq:gensol2}, Sec.~\ref{sec:gen}, where we discuss the technical details.

Using the boundary condition $\mathbf{Q}(0)=\mathbf{0}$, the matrix elements $\beta_{jk}$ can be determined by substituting $Q_{j}^{\beta}$ into $\mathcal{W}$ and letting $\partial\mathcal{W}/\partial\beta_{jk}=0$. The final result is a linear system of dimension $N^{2}$, which can be solved by inverting the matrix of coefficients (See Sec.~\ref{sec:gen}). Once  $\beta_{jk}$ is known, the optimal protocol follows from the relation
\begin{equation}
\lambda_{i} = \sum_{j=1}^{N}M_{ij}Q_{j}^{\beta_{\mathrm{opt}}} + \dot{Q}_{i}^{\beta_{\mathrm{opt}}}
\label{eq:multiproto}
\end{equation}

Once the first iterative step has been done, the new equilibrium is given by $\mathbf{q}^{*}(\boldsymbol{\varepsilon})$ with $\boldsymbol{\varepsilon} = \boldsymbol{\lambda}(t=\tau)$.  Starting from the new equilibrium,  the procedure can be iterated in a new time--interval, say $[\tau,2\tau]$, as long as the largest real part of the eigenvalues of the stability matrix $M_{ij}(\boldsymbol{\varepsilon})$ is strictly negative, with the new equilibrium $\mathbf{q}^{*}(\boldsymbol{\varepsilon})$ playing the role of the original equilibrium in the first iterative step.

%****************************
\section{Special cases}
\label{sec:spec}
%****************************
For pedagogical purposes, we first evaluate the above formulas for weakly interacting systems ($M_{ii}\gg M_{ij}$, $i\neq j$).  We will then generalize these results to systems for which the interactions need not be weak. $M_{ii}\gg M_{ij}$ means that self--interactions are more frequent than intra--species ones. This assumption greatly simplifies calculations, while also allowing us to introduce the essential features of our strategy, which we also use to solve the more general case where the interactions are not necessarily weak. 

When $M_{ii}\gg M_{ij}$ for all $i\neq j$,  Eq.~\eqref{eq:ODEs} simplifies to
\begin{equation}
a^{i}\ddot{Q}_{i} - \nu^{i} Q_{i}=0
\label{eq:singspec}
\end{equation} 

A real solution of Eq.~\eqref{eq:singspec} can be obtained only when $\nu^{i}/a^{i} > 0$ and is given by
\begin{equation}
\label{eq:qeqn} 
	Q^{\beta}_{i}(t)= 2 \beta_{i} \sinh{\left(r_{i}t\right)} 
\end{equation}
where  $Q_{i}^{\beta}(0) = 0$, $r_{i}=\sqrt{\nu^{i}/a^{i}}$, and $\beta_{i}$ is a constant, whose optimal value is to be determined by differentiating
$$
\mathcal{W}=\sum_{i=1}^{N}\left\{\gamma^{i}\Bigl[Q_{i}\dot{Q}_{i}\Bigr]_{0}^{\tau}+\frac{\eta^{i}}{2}\Bigl[\dot{Q}_{i}^{2}\Bigr]_{0}^{\tau} +\right. 
$$
\begin{equation}
\left.+M_{ii}\int_{0}^{\tau}dt\mbox{ }
(a^{i}\dot{Q}_{i}^{2} + \nu^{i}Q_{i}^{2})\right\}
\label{eq:wsing}
\end{equation}
with respect to $\beta_i$, and setting it to zero.

 Next, using the relations 
$\lambda_{i} = \dot{Q} + \sum_{j}M_{ij}Q_{ij}$, $\dot{\lambda}_{i} = \ddot{Q} + \sum_{j}M_{ij}\dot{Q}_{ij}$ and Eq.~\eqref{eq:qeqn}, the single terms in Eq.~\eqref{eq:wsing} take on the form
$$
\gamma^{i}\Bigl[Q_{i}\dot{Q}_{i}\Bigr]_{0}^{\tau}=\gamma^{i}\Bigl[Q_{i}^{*}(\varepsilon_{i})\Bigl(-2M_{ii}\beta_{i}\sinh{(r_{i}\tau)}+\varepsilon_{i}\Bigr)\Bigr]
$$
$$
\frac{\eta^{i}}{2}\Bigl[\dot{Q}_{i}^{2}\Bigr]_{0}^{\tau}=\frac{\eta^{i}}{2}\Bigl[\Bigl(-2M_{ii}\beta_{i}\sinh{(r_{i}\tau)}+\varepsilon_{i}\Bigr)^{2}\Bigr]
$$
\begin{equation*}
M_{ii}  \int_{0}^{\tau}dt\mbox{ }
(a^{i}\dot{Q}_{i}^{2} + \nu^{i}Q_{i}^{2})  = 2M_{ii}\beta_{i}^{2}a^{i}r_{i}\sinh{(2r_{i}\tau)}
\end{equation*}
We now set $\partial\mathcal{W}/\partial\beta_{i} = 0$ to obtain the optimal value $\beta_i=\beta_{i}^{\mathrm{opt}}$, given by
\begin{equation*}
\beta^{\mathrm{opt}}_{i}  = \frac{\varepsilon_{i}
+\frac{\gamma^{i}}{\eta^{i}}q_{i}^{*}(\varepsilon_{i})}
{4\frac{a^{i}}{\eta^{i}}r_{i}\cosh{(r_{i}\tau)}+2M_{ii}\sinh{(r_{i}\tau)}}
\end{equation*}
From the above solution we can derive the optimal protocol explicitly:
\begin{align}
\lambda^{\mathrm{opt}}_{i}(t) &= \dot{Q}^{\beta_{\mathrm{opt}}}_{i}(t) + M_{ii} Q^{\beta_{\mathrm{opt}}}_{i}(t)\label{eq:optl}\\
&=\frac{r_{i}+M_{ii}\tanh{(r_{i} t)}}{2\frac{a^{i}}{\eta^{i}}r_{i}+M_{ii}\tanh{(r_{i}\tau)}}\left(\varepsilon_{i}+\frac{\gamma^{i}}{\eta^{i}}q_{i}^{*}(\varepsilon_{i})\right) 
\nonumber
\end{align}
This formula defines the most general protocol to induce an optimal infinitesimal displacement for system characterized by either a single degree of freedom or weakly interacting systems. It works for any non--linear system, provided that the iterative procedure can be applied and for the general class of cost functionals given in Eq.~\eqref{eq:lagrapp}.

%*************************************************************
\subsection{Design of Universal controllers.}
\label{sec:univ}
%*************************************************************
%_____________________FIGURE-1_____________
\begin{figure}[htbp]
\centering
\includegraphics[clip=true,keepaspectratio,width=1.0\columnwidth]{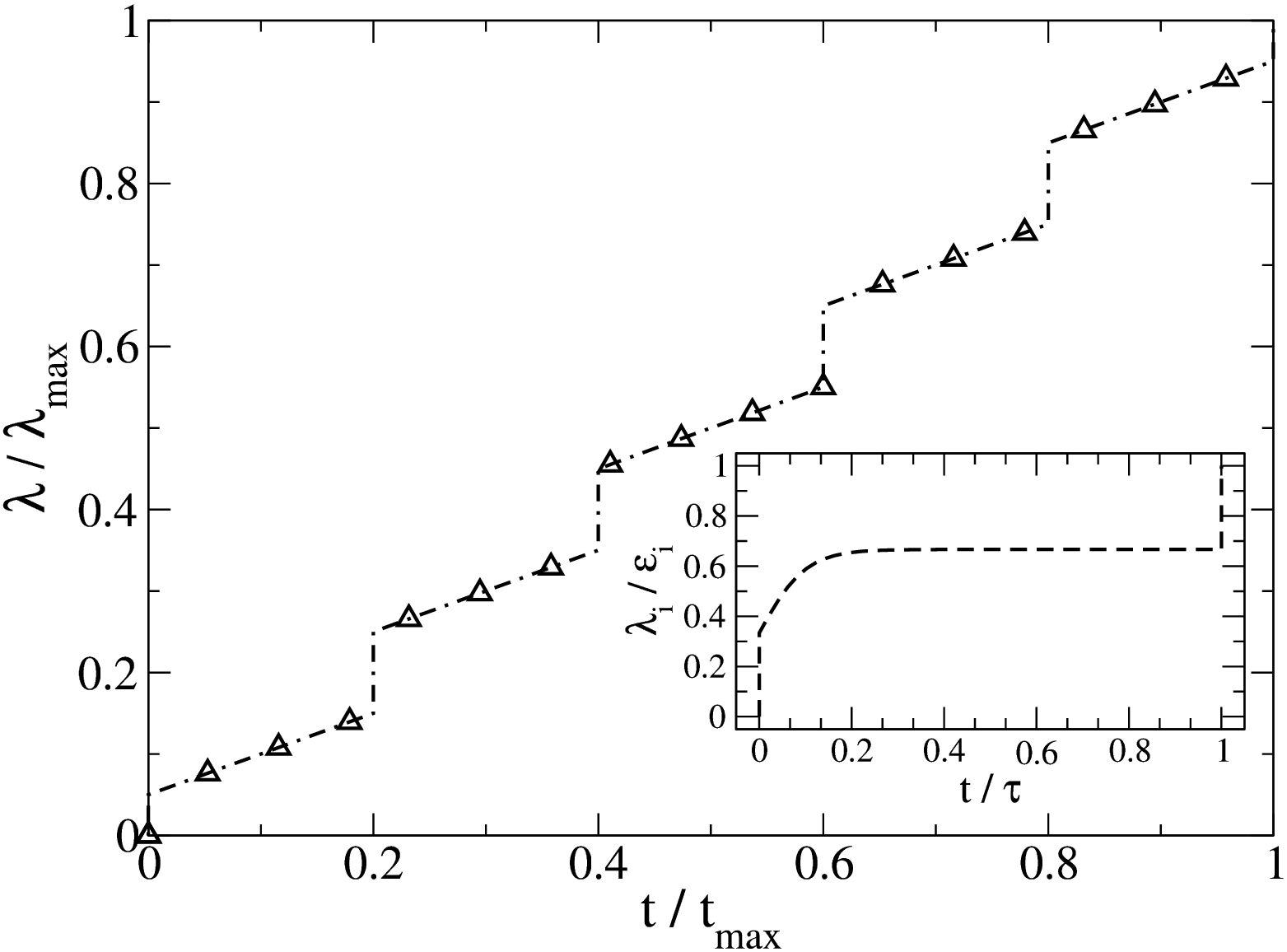}% 
\caption{\textbf{(main panel)}The iteration of the universal formula in Eq.~\eqref{eq:approx} with the local slopes $M_{ii}(\lambda)$ evaluated along the stable branch of a simple quadratic system defined by $F(q) = \mu q - A q^{2}$ $(\mu=2, A=1)$ (cfr.~See also top panel in Fig.~\ref{fig:quadratiter} ). The iterations start at the original equilibrium $\mu/A$ and bring the system to the point $\mu/A + \mu/2A$,  changing the protocol by a quantity $\varepsilon=10^{-1}\mu/A$ at each iteration; \textbf{(inset)} Eq.~\eqref{eq:optl} referred to a single iteration. Here we show the non--universal behavior for $\gamma^{i}=\alpha^{i}=0$. The iteration in this case cannot be drawn without knowing all the details of the Lagrangian, i.~e. the constants $r_{i}$ at any iterative step. The plot refers to the special case with $r_{i}=M_{ii}=1$ in the first iterative step.} 
\label{fig:protocol}
\end{figure}
%_____________________________________________
The general solution \eqref{eq:optl} reveals an interesting class of Lagrangians for which the optimal control protocol becomes independent on the parameters defining the cost functional. Specifically, if $\mathcal{L}$ is such that $\alpha^{i} = \gamma^{i}=0$, 
Eq.~\eqref{eq:ai} gives $a^{i}/\eta^{i}= 1$ in Eq.~\eqref{eq:optl}, and when $r_{i}\tau\ll 1$, Eq.~\eqref{eq:optl} looses all constants associated with the Lagrangian, and we get a universal protocol
\begin{equation}
\lambda^{\mathrm{opt}}_{i}(t) \approx \frac{\varepsilon_{i} (1 + M_{ii} t)}{(2 + M_{ii}\tau)}
\label{eq:approx}
\end{equation}
Note that the condition $r_{i}\tau\ll 1$ giving rise to the universal protocol can always be satisfied by choosing a sufficiently small $\tau$. Picking a sufficiently--small $\tau$ is also important to avoid noise--driven transitions to other equilibria. Ideally, $\tau$ must be smaller than the fastest of such transitions.

While \eqref{eq:approx} has been derived by \citet{schmiedl2007} assuming a linear system subject to the minimization of the work functional, here we identify this formula as a special case of a more general result, i.e. our \eqref{eq:optl}. We also identify \eqref{eq:approx} as a universal control strategy that works beyond linear systems and specific work / cost functionals, provided that the Lagrangian belongs to the particular class identified by \eqref{eq:g} and ~\eqref{eq:h} with $\alpha^{i} = \gamma^{i} =0$. In Sec.~\ref{sec:univex} we will discuss some specific examples of Lagrangians belonging to such a universal class.

Once the system has been brought to the new equilibrium point $q_{i}^{*}(\varepsilon_{i})$, this point can now be considered as the new original equilibrium, and the procedure can be repeated again, provided that all the parameters are updated by expanding, this time, around $q^{*}_{i}(\varepsilon_{i})$. For example, Fig.~\ref{fig:protocol} (\textbf{main panel}) shows the iteration of the map in Eq.~\eqref{eq:approx} for a quadratic system, i.~e. the universal behavior, from $\lambda=0$ to $\lambda=\lambda_{\mathrm{max}}=K\varepsilon$ ($K=5$), each segment evaluated in a time--interval of the form $[(n-1)\tau,n\tau]$ $(n=1,\cdots,K)$. The slope of every segment depends on $M_{ii}((n-1)\varepsilon_{i})$, which must be evaluated along the function $q^{*}_{i}(\lambda)$.

The control function Eq.~\eqref{eq:optl} in the non--universal regime is shown in Fig.~\ref{fig:protocol} (\textbf{inset})  for a single iteration starting from the natural equilibrium. More iterations of the formula are not shown since we would need all the details of the cost functional. 
For this case, we took $\gamma^{i} = \alpha^{i} = 0$ and $r_{i}=M_{ii} = 1$.

In passing, we note that Eq.~\eqref{eq:approx} and Eq.~\eqref{eq:optl} are characterized by  discontinuities at endpoints, i.e. at $t\to 0^{-}$ and $t\to\tau^{+}$. We recall that such discontinuities commonly appear in solutions of optimal control problems \cite{band1982, schmiedl2007,gomez2008}. 

\section{General case}
\label{sec:gen}

When $M_{ii}\gg M_{ij}$ does not hold, the general solution to the second order system in Eq.~\eqref{eq:ODEs} can be written as
$$
Q_{j} = \sum_{k}\biggl[\beta_{jk}\exp{(\mbox{i}\sqrt{\omega_{k}}t)}+\beta'_{jk}\exp{(-\mbox{i}\sqrt{\omega_{k}}t)}\biggr] + \mbox{c.~c.}
$$
where $c.~c.$ indicates the complex conjugate of the first term ($\mathbf{Q}$ must be real) and $\omega_{k}^{-1}$ is the (in general, complex) eigenvalues of the matrix $\hat{B}^{-1}\hat{A}$ (see Eq.~\eqref{eq:A0} and Eq.~\eqref{eq:B0} for the definition of the matrices $\hat{B}$ and $\hat{A}$). The condition $\mathbf{Q}(\mathbf{0}) = \mathbf{0}$ requires $\beta_{jk} = -\beta'_{jk}$ and thus we can write the final solution in real form as:
\begin{eqnarray}
Q_{j} & = & \sum_{k}\beta_{jk}\phi_{k}(t)\label{eq:gensol1}\\
\phi_{k}(t) & = & 2\sinh{(\omega^{\mathrm{(Re)}}_{k}t) }\cos{(\omega^{\mathrm{(Im)}}_{k}t)}\label{eq:gensol2} 
\end{eqnarray}
with $\mbox{i}\sqrt{\omega_{k}} = \omega^{\mathrm{(Re)}}_{k} + \mbox{i}\omega^{\mathrm{(Im)}}_{k}$ ($\mbox{i}^{2}=-1$) and $\beta_{jk}\in\mathbb{R}$. 

The next step is the minimization of the cost functional in Eq.~\eqref{eq:WFIN}. Using Eq.~\eqref{eq:gensol1} and remembering that, around the equilibrium we have
$$
\dot{Q}_{i} = \sum_{j}M_{ij}Q_{j} + \lambda_{i}(t)=\sum_{jk}M_{ij}\beta_{jk}\phi_{k} + \lambda_{i}(t)
$$
the minimization of ~\eqref{eq:WFIN} is done by setting $\partial \mathcal{W}/\partial \beta_{pp'}=0$. The calculation is straightforward, however, tedious. Here, we report the final result:

$$
\sum_{i}\Biggl\{-\Bigl[\gamma^{i}Q_{i}^{*}(\tau)+\eta^{i}\varepsilon_{i}\Bigr]M_{ip}\phi_{p'}(\tau) +
$$
$$
+ \sum_{jk}\Biggl[M_{ij}\biggl(\eta^{i}M_{ip}\phi_{k}\phi_{p'}+\delta_{ip}\Bigl(\tilde{\Gamma}^{i}_{kp'}+\tilde{\Gamma}^{j}_{kp'}
$$
\begin{equation}
+\delta_{ij}(\Gamma^{i}_{kp'}-\tilde{\Gamma}^{i}_{kp'})+\delta_{ij}(\Gamma^{j}_{kp'}-\tilde{\Gamma}^{j}_{kp'})\Bigr)\biggr)\Biggr]\beta_{jk}\Biggr\} = 0
\label{eq:betas}
\end{equation}
where $\Gamma^{i}_{kk'}$ and $\tilde{\Gamma}^{i}_{kk'}$ are defined as
\begin{eqnarray}
\Gamma^{i}_{kk'} & = & a^{i}\int_{0}^{\tau}dt\mbox{ }\dot{\phi}_{k}\dot{\phi}_{k'}+\nu^{i}\int_{0}^{\tau}dt\mbox{ }\phi_{k}\phi_{k'}\label{eq:gamma}\\
\tilde{\Gamma}^{i}_{kk'} & = & \eta^{i}\int_{0}^{\tau}dt\mbox{ }\dot{\phi}_{k}\dot{\phi}_{k'}+\nu^{i}\int_{0}^{\tau}dt\mbox{ }\phi_{k}\phi_{k'}\label{eq:til}
\end{eqnarray}

\eqref{eq:betas} is a linear system of equations with unknowns $\beta_{jk}$. Summing over $i$ and relabeling the indices such that $jk \to q' = 1,2,\cdots,N^{2}$ and $pp' \to q = 1,2,\cdots,N^{2}$, the linear system becomes
$$
C_{q} + \sum_{q'}T_{qq'}\beta_{q'}=0
$$
where $C_{q}$ and $T_{qq'}$ consists of known constants, given with the problem. The solution of the above system can be found if and only if $\mbox{det}(T_{qq'})\ne 0$ and can be written as 
\begin{equation}
\boldsymbol{\beta} \equiv\boldsymbol{\beta}_{\mathrm{opt}}= \hat{T}^{-1}\mathbf{C}
\label{eq:inverse}
\end{equation}
which allows us to find the protocol using equation Eq.~\eqref{eq:multiproto} and Eq.~\eqref{eq:gensol1} evaluated for $\boldsymbol{\beta}=\boldsymbol{\beta}_{opt}$

Just as in Eq.~\eqref{eq:approx}, when $\gamma^{i} = \alpha^{i} = 0$ (i.~e. $\tilde{\Gamma}^{i}_{kk'}=\Gamma^{i}_{kk'}$), it is possible to find an explicit expression for the protocol in the limit of $\tau\to 0$. In this case we explicitly get
\begin{equation}
\boldsymbol{\lambda}(t)=(t\hat{M}+\hat{I})\biggl(\hat{I}+\tau\hat{\mathcal{G}}^{-1}\hat{M}\biggr)^{-1}\hat{\mathcal{G}}^{-1}\boldsymbol{\varepsilon}
\label{eq:mildu}
\end{equation}
where the matrix $\hat{\mathcal{G}}$ is given by
$$
\mathcal{G}_{ij} = \frac{M_{ij}}{M_{ii}}\left(1+\frac{\eta^{j}}{\eta^{i}}\right) 
$$
As wee see, contrary to the weakly interacting case we do not necessarily obtain a universal protocol, since the control function now depends on the ratio of the $\eta$'s (but not on the eigenvalues $\omega_{k}$). Nevertheless, a ``milder'' universality is recovered when the $\eta$'s are similar to one other,
$$
\left|\frac{\eta^{j}}{\eta^{i}}\right|\sim \mathcal{O}(1)
$$
This condition should not be viewed as an unlikely conspiracy of parameters. In a typical thermodynamic system, all particles will have the same physical properties, and any cost functional that treats all particles on equal footing (e.g. physical work) will fit the bill. In such cases the control function will be universal, in the sense that it will not depend on \emph{what} the physical properties of the particles are, as long as the physical properties of all particles are the same $\eta^i=\eta$.

%*************************************************************
\section{Examples and limits of application.}
\label{sec:disc}
%*************************************************************
We will now establish the domain of applicability of our framework and discuss its limitations. We will do so by picking a number of specific examples of dynamic systems and cost functionals, and illustrate the workings or limitations of our formulas. 

We will start with two one--dimensional systems: the logistic model, and a periodic potential model. We will then move on to a multi--species generalization of a Lotka--Volterra ecological model. Finally, we will discuss a number of specific cost functionals, motivated by mechanical and thermodynamic applications.

%_______________________FIGURE-2___________________
\begin{figure}[htbp]
\centering
\includegraphics[clip=true,keepaspectratio,width=1.0\columnwidth]{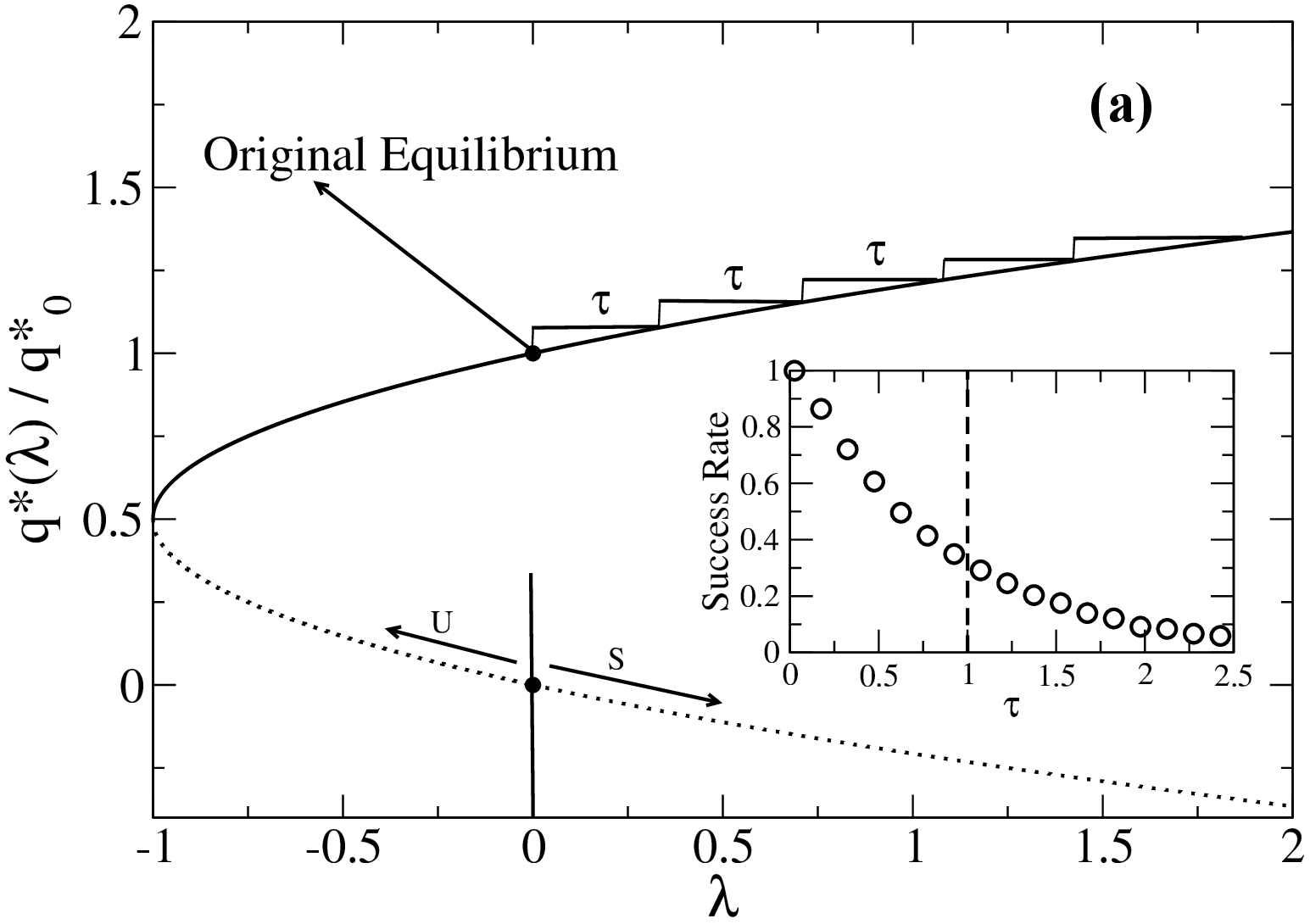}% 
\qquad \includegraphics[clip=true,keepaspectratio,width=1.0\columnwidth]{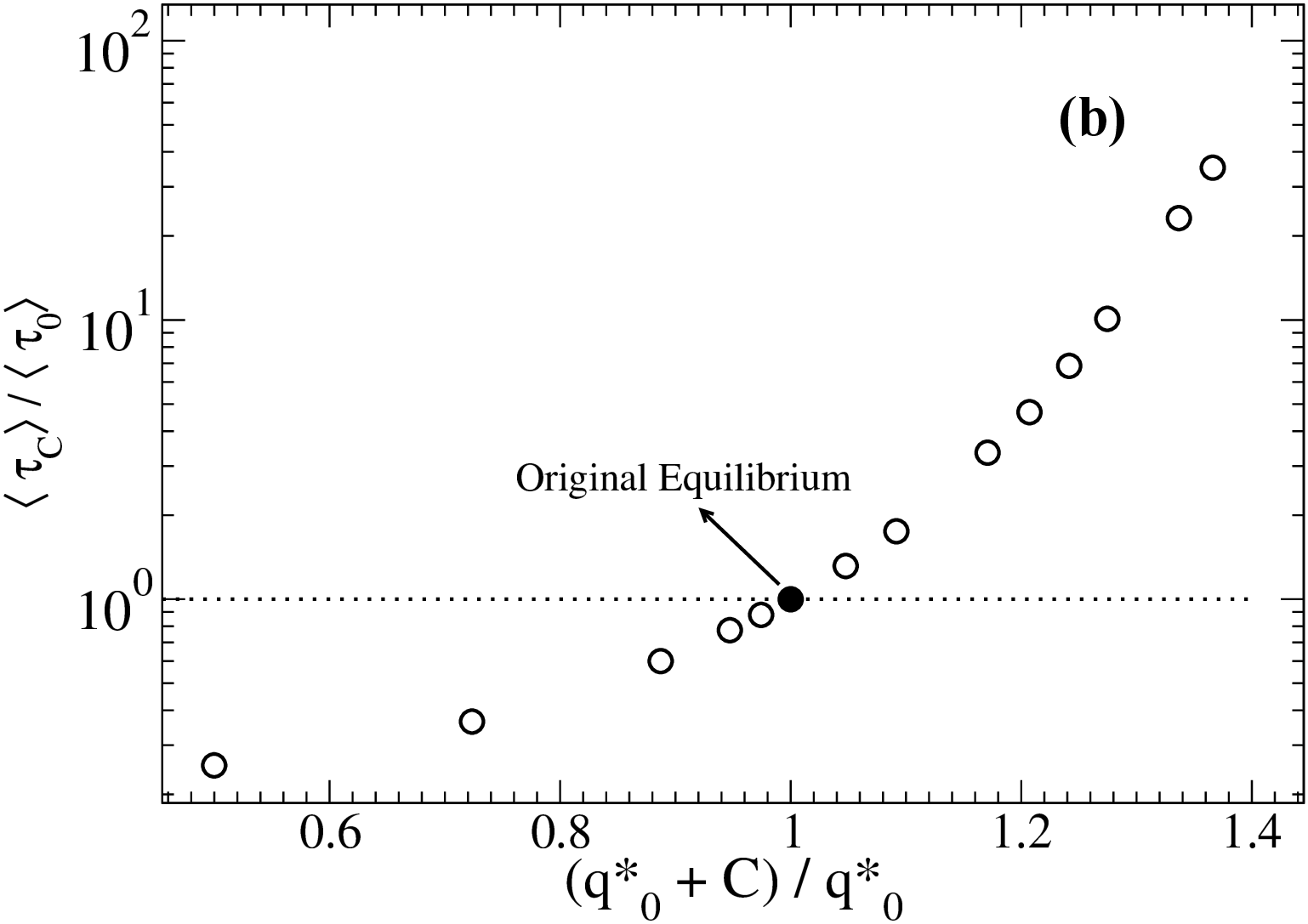}
\caption{\textbf{ (a), Main Panel:} the function $q^{*}(\lambda)$ for the quadratic system defined by $F(q) = \mu q - A q^{2}$ ($\mu = 2, A = 1$). After applying the control protocol in a  finite time 
$\tau$, the system sets on the new equilibrium point $q^{*}(\lambda)$. From here, we can linearize again
and iterate the single step procedure; the bottom branch is both unstable (U) and stable (S). \textbf{(a), Inset:} the success rate of the control strategy, i.~e. the number of trajectories which reach the absorbing boundary while applying the first iteration, from the natural equilibrium to a new equilibrium point infinitesimally close to it; \textbf{(b)} To any iteration corresponds a final equilibrium value $q^{*}_{0}+C$ and an average extinction time $\langle\tau_{C}\rangle$, which, once compared with $\langle\tau_{0}\rangle$ results highly improved, even for small displacements from the natural equilibrium}
\label{fig:quadratiter}
\end{figure}
%__________________________________________________________

\subsection{One--dimensional systems.}
\label{sec:}

An important limitation of our framework is that 
$q^{*}(\lambda)$ must exist, remain stable for the range of $\lambda$ in question, and be a continuous function. Furthermore, if the system has multiple stable points, or if the stable path $q^{*}(\lambda)$ bifurcates, then noise can trigger undesirable transitions to other equilibria.

To exemplify these limitations, we discuss a specific system which has multiple stable points, one of which, assumed to be absorbing, 
$F(q) = \mu q - A q^{2}$, $\mu/A>0$. Suppose we originally start from the stable point $q_{0}^{*} = \mu/A$. As we add in the control $\lambda$, both stable points shift (Fig.~\ref{fig:quadratiter}\textbf{(a), main panel}). As long as rare events are ignored, our procedure can move the system to any arbitrary state $q>q^{*}_{0}/2$; however, it is essential that as we move along the upper branch, $q^{*}(\lambda)$, noise does not transition the system to the lower (absorbing) branch.

With rare events, the success of our protocol depends on the choice of $\tau$, and can be evaluated. We simulate many trajectories moving from $\mu/A$ to $\mu/A + \delta(\mu/A)$ (where $\delta \ll 1$) within a time $\tau$, and count the fraction of trajectories that crash to the lower branch. The result is shown in Fig.~\ref{fig:quadratiter}\textbf{(a), inset}. As we see, for $\tau\ll 1$, probability of success approaches 1. Furthermore, and regardless of $\tau$, as we iterate along $q^{*}(\lambda)$ the likelihood of success keeps increasing, since the further we move away from the absorbing state, the more the expected crash time increases (cf. Fig.~\ref{fig:quadratiter}(\textbf{b})).

We point out that whenever $\alpha^{i}=\gamma^{i}=0$, having a high success rate is equivalent to the applicability  of the universal controller. This is because success and ``universality'' both hinge on $r\tau\ll1$.

Of course, we cannot control rare events using the iterative technique developed here, because in the long--time, a rare jump will always occur, bringing the system to the absorbing boundary (if there is one). More refined techniques, based on the evaluation of the large deviation function \citep{touchette2009} should be applied to control rare events. These techniques are based on  path integral description of the stochastic dynamics \citep{kappen2005} and/or the evaluation of the main contribution to the path integral (WKB method) \citep{assaf2010,khasin2011,khasin2010s,hindes2016}.  

The technique discussed in this letter works perfectly in the deterministic limit. In the case of stochastic systems, the best we can do is to chose a $\tau$ that maximizes the success rate in a given time window.  
%____________________FIGURE-3________________________
\begin{figure}[htbp]
\centering
\includegraphics[clip=true,keepaspectratio,width=1.0\columnwidth]{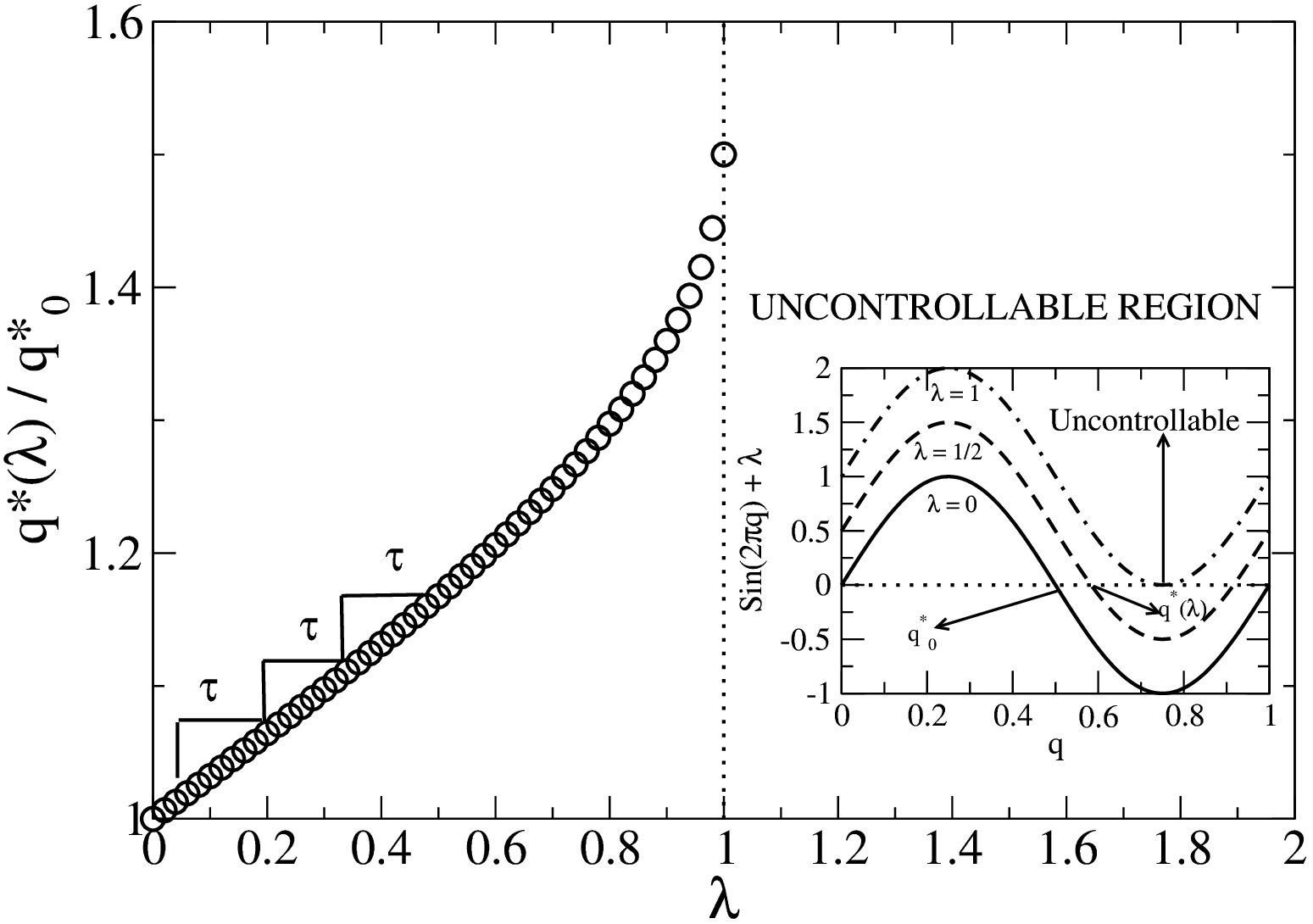}% "%" necessario
\caption{\textbf{(Main panel)} the function $q^{*}(\lambda)$ for the multi--stable system defined by $F(q) = \sin{(2\pi q)}$. Starting from the original equilibrium $q^{*}_{0}=1/2$ we can again apply the iterative procedure moving along the line $q^{*}(\lambda)$. However, we immediately realize that for $\lambda\ge 1$ the system is always out of equilibrium and the proposed technique fails; \textbf{(inset)} the function $F(q) = \sin{(2\pi q)} + \lambda$ for different values of $\lambda$.}
\label{fig:multicontr}
\end{figure}
%_______________________________________________________
Now, we investigate another multistable system, that has a periodic potential, $F(q)=\sin{(2\pi q)}=-\partial_{q}U(q)$ with $U(q) = \cos{(2\pi q)/2\pi}$. 
We assume that the system starts from $q^{*}_{0} = 1/2$ (Fig.~\ref{fig:multicontr} (\textbf{inset})). In the limiting case of vanishing noise, we can safely iterate the control procedure along the geometric locus defined by  $q^{*}(\lambda)$, as shown in Fig.~\ref{fig:multicontr} (\textbf{main panel}). However we see that the technique fails at $|\lambda|\ge 1$, even for deterministic systems, since there comes a point where the system looses all stable points, and we cannot proceed further with our method. For such cases, a non--equilibrium procedure is in order. When we cannot neglect the influence of the noise, we should take into account that the jumps between stable equilibria occur at a rate $\sim\exp{(U(q=1))}$ \citep{hanggi1990}. Thus, increasing $\lambda$ has the effect of increasing the transition frequency by a factor of $\exp{(\lambda)}$. If our goal is to trigger a jump to another state, we can safely employ the optimal procedure until we reach a value of $\lambda$ that gives us the desired jumping rate. This procedure can also be used to inhibit the jumps between two stable equilibria. In this case, rather than moving $\lambda$ from $0$ to $\lambda_{f}<1$, we must move $\lambda$ from $0$ to $\lambda_{f}>-1$, which would inhibit the jumping frequency by a factor $\exp{(-|\lambda_{f}|)}$. In both cases, we can ensure a high success rate while iterating the protocol, by picking up a suitable $\tau$, as discussed before.

%*************************************************************
\subsection{The Lotka--Volterra model.}
%*************************************************************
In this section we discuss a successful  application of our procedure to the multi--species generalization of the Lotka--Volterra model defined by
\begin{equation}
\frac{dq_{i}}{dt} = \mu_{i}q_{i} + \sum_{j=1}^{N}J_{ij}q_{j}q_{i}+\lambda_{i}(t)\qquad i=1,\ldots,N. 
\label{eq:LVM}
\end{equation}
Here $N$ is the number of species, $\mu_i$ is the intrinsic growth rate of species $i$ and $J_{ij}$ quantifies the interactions between species $i$ and $j$. Eq.~\eqref{eq:LVM} has a single internal fixed point $\mathbf{q}^{*}_{0}$ for $\boldsymbol{\lambda} = \mathbf{0}$, given by
\begin{equation}
\mathbf{q}^{*}_{0} = -\hat{J}^{-1}\boldsymbol{\mu}
\label{eq:fixLV}
\end{equation}

Here we will evaluate our formulas for a competitive model ($J_{ij}\propto -J_{ji}$ and $J_{ii}<0$) with $N=10$ species. In this case, the equilibrium point in Eq.~\eqref{eq:fixLV} is globally stable \citep{allesina2012}.
We draw the parameters $J_{ij}$ from a Gaussian distribution with zero mean and unit variance, and choose the original equilibrium point $\mathbf{q}^{*}_{0}$ at random, such that $q^{*}_{0,i}$ is uniformly distributed between $0$ and $1$. The parameters $\mu_{i}$ are fixed by inverting the relation in Eq.~\eqref{eq:fixLV}.

\begin{figure}[htbp]
\centering
\includegraphics[clip=true,keepaspectratio,width=1.0\columnwidth]{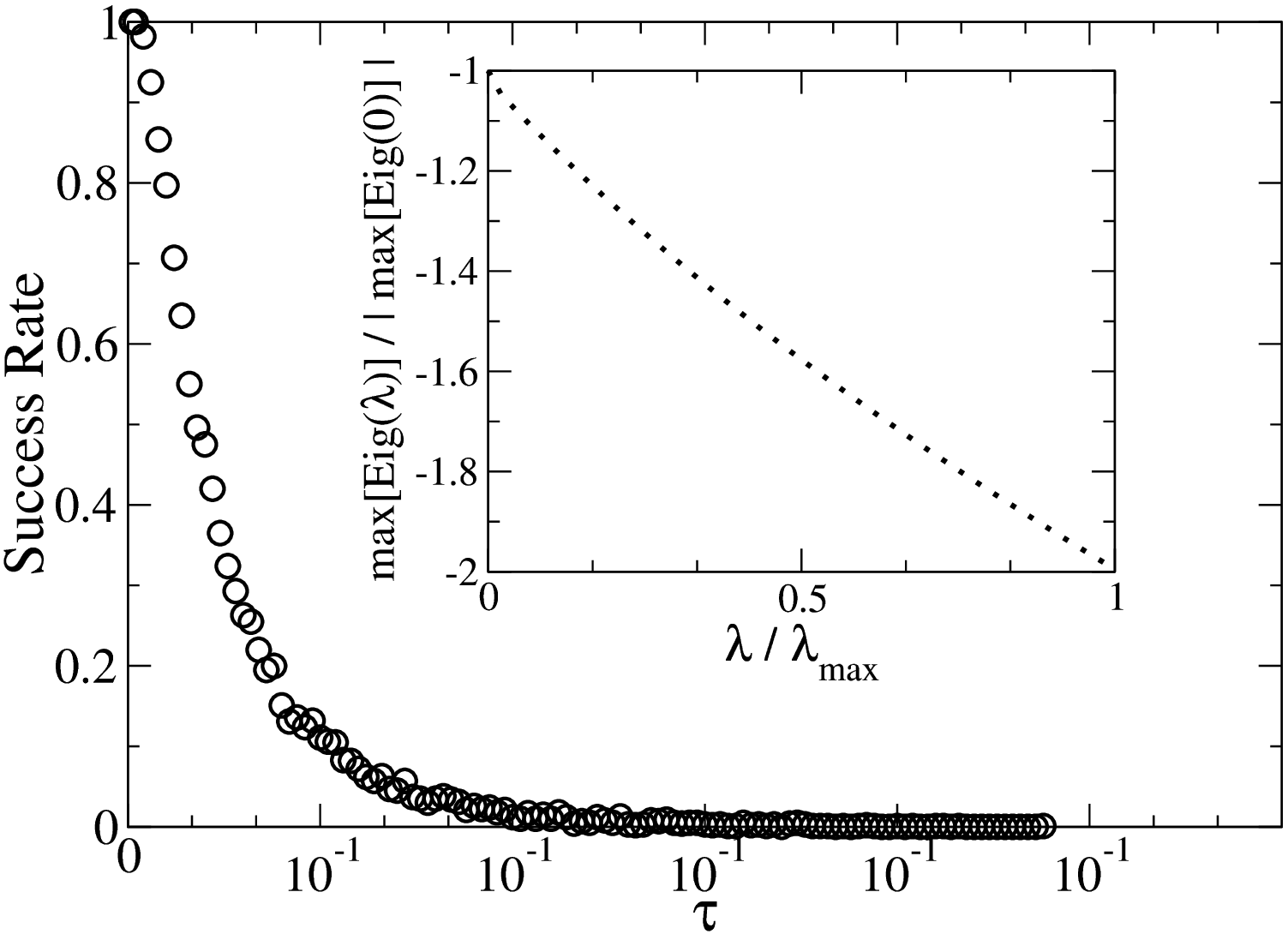}% "%" necessario
\caption{\textbf{(Main panel)} Success rate of the $10$--dimensional Lotka--Volterra system; \textbf{(inset)} largest real part of the stability matrix as a function of $\lambda$.}
\label{fig:Ext}
\end{figure}

The starting point of our technique is the displacement of the system from the original equilibrium to a new one $\mathbf{q}^{*}(\boldsymbol{\lambda})$, thanks to the application of an external controller $\boldsymbol{\lambda}$, such that $\boldsymbol{\lambda}(0)=\mathbf{0}$ and $\boldsymbol{\lambda}(\tau) = \boldsymbol{\varepsilon}$, with $|\boldsymbol{\varepsilon}|\ll 1$. Thus, we first need to solve the equation 
$$
\mu_{i}q_{i} + \sum_{j=1}^{N}J_{ij}q_{j}q_{i}+\lambda_{i} = 0
$$ 
and check if the largest real part of the stability matrix $\hat{M}(\boldsymbol{\lambda})$, evaluated at $\mathbf{q}^{*}(\boldsymbol{\lambda})$, is strictly negative. If so, we can safely apply our iterative technique to this non--linear and multidimensional problem. In Fig.~\ref{fig:Ext} (inset) we show the largest real part of $\hat{M}(\boldsymbol{\lambda})$ evaluated numerically as a function of the magnitude $\lambda = |\boldsymbol{\lambda}|$, when $\boldsymbol{\lambda}=\lambda (1,1,1,\cdots,1)$.    
%_____________________FIGURE-4_________________
%________________________________________________
 In particular $\lambda = K\varepsilon$ with $\varepsilon\ll 1$ and $K$ the iterative step. 
 
As we did in the single--species case, we  evaluate the success rate of our technique for this multi--dimensional system. Indeed, as long as we move along the line in Fig.~\ref{fig:Ext} (inset) we can safely employ our technique in the deterministic case. However, when the noise is large enough to destabilize the globally--stable point $\mathbf{q}^{*}(\boldsymbol{\lambda})$ we must be cautious about the success rate of our method. In particular, using Eq.~\eqref{eq:mildu}, we displace the Lotka--Volterra system from $\mathbf{q}^{*}_{0}$ to a near-by point $\mathbf{q}^{*}(\boldsymbol{\lambda}_{f})$, with $\boldsymbol{\lambda}_{f}=\varepsilon (1,1,1,\cdots,1)$ and $\varepsilon=10^{-3}$. We repeat the process $N_{e}=1000$ times in a fixed time--interval $[0,\tau]$ and  count how many of the $N_{e}$ total realizations are such that at least one of the ten species in the Lotka--Volterra system goes extinct.  The success rate is shown in the main panel of Fig.~\ref{fig:Ext} and, as in the single--species case, approaches one as $\tau\to 0$, which justifies the limit in Eq.~\eqref{eq:mildu}.

\subsection{Explicit examples of cost--functionals.}
\label{sec:univex}

The physical interpretation of the universal class of Lagrangians can be understood better with few explicit examples. Following \cite{schmiedl2007}, we consider a mechanical system evolving according to the equation
$$
\frac{dq}{dt}=-\frac{\partial V(q,\lambda)}{\partial q} +\sqrt{2}\xi
$$
where 
$$
V(q,\lambda) = \frac{1}{2}(q + \lambda)^{2}
$$
We thus have (averaging over the noise) $\dot{q} = - q + \lambda$. The work $\mathcal{W}$ performed by $\lambda$ on the system is,
\begin{equation}
\mathcal{W} = \int_{0}^{\tau}dt\dot{\lambda}\left\langle\frac{\partial V}{\partial \lambda}\right\rangle = \int_{0}^{\tau}dt \dot{\lambda}\dot{q}
\label{eq:ex1}
\end{equation}
giving a Lagrangian of the form ~\eqref{eq:lagrapp} with $\eta=1$ and all the other parameters equal to zero. The Lagrangian in this case is not approximate and belongs to the universal class. We explicitly observe that in Eq.~\eqref{eq:optl} $r^{i} = \sqrt{\nu^{i}/ a^{i}}$ goes to zero as $\sim\sqrt{\nu^{i}}$ when all the parameters in Eq.~\eqref{eq:lagrapp} are zero with the exception of the $\eta$'s. Thus, in order to get the universal behavior, we do not even need to take the limit $\tau\to 0$. It will be enough to take the limit for $\nu^{i}\to 0$, which again gives Eq.~\eqref{eq:approx}. The same conclusion holds more generally, for the potential
\begin{equation}
V(\mathbf{q},\boldsymbol{\lambda})=\frac{1}{2}(\mathbf{q}-\boldsymbol{\lambda})\cdot (\mathbf{q}-\boldsymbol{\lambda})
\implies\mathcal{L}=\dot{\mathbf{q}}\cdot\dot{\boldsymbol{\lambda}}.
\label{eq:vmul}
\end{equation}
Another example of Lagrangian belonging to the universal class is 
\begin{equation}
\mathcal{L}=\frac{1}{2}(\dot{q} - \dot{\lambda})^{2}
\label{eq:kinlagr}
\end{equation}
In a mechanical framework, this might be interpreted as a minimization of the effect of friction--like forces between the system and the controlling agent. 

The linearized form (Eq.~\eqref{eq:lagrapp}) of the above Lagrangian is
$$
\mathcal{L}\approx - \dot{q}\dot{\lambda}+\cdots
$$
which belongs to the universal class with $\eta^{i}=-1$. The universal limit follows from Eq.~\eqref{eq:optl} taking $\nu^{i}\to 0$ and it is valid for any $\tau$.
More generally, we can consider a Lagrangian of the form 
$$
\mathcal{L}=\frac{1}{2}(q-\lambda)^{2} + \frac{1}{2}(\dot{q}-\dot{\lambda})^{2}\approx -q\lambda - \dot{q}\dot{\lambda}+\cdots
$$
which also falls in the universal class with $\nu^{i} = \eta^{i} = -1$. In this case and for Eq.~\eqref{eq:kinlagr} the universal limit  follows only if we pick up a sufficiently small $\tau$, such that $r^{i}\tau \ll 1$. 
A multi--dimensional generalization of the above Lagrangian and Eq.~\eqref{eq:kinlagr} are easily found to belong to the universal class, when $\eta^{i} = -1$ for all $i$.

An example of Lagrangian which does not belong to the universal class can be borrowed from the experiments with feedback traps. Such experiments are often employed to investigate the role of information in the energy exchange between physical systems \cite{jun2014}. If we allow for the control of the tilt in a double--well potential, we have
$$
V(q,\lambda)=-\frac{1}{2}q^{2} + \frac{1}{4}q^{4} - \lambda(t)q
$$
i.~e.
$$
\dot{q} = q - q^{3} + \lambda(t) + \mbox{noise}
$$
and the Lagrangian is given by
$$
\mathcal{L} =\dot{\lambda}\langle\partial_{\lambda}V\rangle =  -q\dot{\lambda}
$$
which is of the form in Eq.~\eqref{eq:lagrapp} with all the parameters zero with the exception of $\gamma^{i}$. We observe that in this case a divergence appears in Eq.~\eqref{eq:optl}, since $\eta^{i}\to0$ and our method cannot be applied. However, for general Lagrangians, such divergence is removed if we consider higher order terms in Eq.~\eqref{eq:lagrapp}. For example, in the case of systems with $M_{ij}\ll M_{ii}$,
if a non-linear term of the form
$$
\sum_{i}\tilde{\eta}^{i}\lambda_{i}\dot{\lambda}_{i}
$$
is added to the Lagrangian in Eq.~\eqref{eq:lagrapp}, the divergence is removed. In the presence of this addition, Eq.~\eqref{eq:optl} still holds true as long as we replace
\begin{eqnarray*}
\eta^{i} & \to & \eta^{i} + \tilde{\eta}^{i}\\
\gamma^{i} & \to & \gamma^{i} + M_{ii}\tilde{\eta}^{i}
\end{eqnarray*}
However, obtaining the effect of other possible non--linear additions is extremely challenging. Furthermore, when $M_{ij}\ll M_{ii}$ does not hold,
we cannot simply get away with a trivial shift in parameters.
We leave a closer investigation of non-linear Lagrangians, and their removal of singularities to a future study.

%*************************************************************
\section{Conclusions}
\label{sec:concl}
%*************************************************************
In this paper we developed an iterative control strategy to optimally solve a certain class of nonlinear problems. The  procedure is based on successive linearizations around a moving equilibrium $\mathbf{q}^{*}(\boldsymbol{\lambda})$ of the system plus a controller, as we push the system with a control function $\boldsymbol{\lambda}$. We observed that, for deterministic systems, the procedure can be implemented as long as the largest real part of the eigenvalues of the stability matrix $M_{ij}(\boldsymbol
{\lambda})$ remains strictly negative. When such a condition does not hold, the procedure fails. 

We have determined the effects of noise in our control framework in terms of success rate. In particular, the control time window $\tau$ must be picked as small as possible to guarantee high success. However, for large strengths of the noise, $\tau$ would become unreasonably small. Thus, our procedure can be safely applied in the limit of weak noise. 

Surprisingly, we found that under certain conditions, it is possible to design universal controllers, thus engineering the control protocol without minding the particular form of the cost--functional.

In closing, we should note that the additive control scheme we proposed here is not the only possible form of control. For example, a protocol of the form $\mu_{i} F_{i}(q_{1},\cdots,q_{N})$ could be considered, for a multiplicative control function $\mu_{i}$. However, such a multiplicative control function cannot displace the equilibrium point. A combination of additive and multiplicative control $\mu_{i} F_{i}(q_{1},\cdots,q_{N}) + \lambda_{i}$ on the other hand, may lead to non-trivial results, and lead to an interesting generalization of our work.

%************************************************************************************
% acknowledgments
\begin{acknowledgments}
We thank Vu Nguyen and John Bechhoefer for insightful discussions. This material is based upon work supported by the Defense Advanced Research Projects Agency under Contract No. HR0011-16-C-0062. 
\end{acknowledgments}
%************************************************************************************

%************************************************************************************
%APPENDX
\appendix
%********************************
\section{Derivation of Eq.~(\ref{eq:WFIN})}
\label{sec:one}
%********************************
The Lagragian expanded around the original equilibrium $\mathbf{q}^{*}_{0}$ is 
$$
\mathcal{L} \approx \mbox{const.} + \sum_{i=1}^{N}(\nu^{i} Q_{i}\lambda_{i} + \alpha^{i} \dot{Q}_{i}\lambda_{i} + \gamma^{i} Q_{i}\dot{\lambda}_{i} + \eta^{i} \dot{Q}_{i}\dot{\lambda}_{i})
$$
From the mean field equation 
$$
\frac{dQ_{i}}{dt} = -\sum_{j}M_{ij}Q_{j} +  \lambda_{i}
$$
it is possible to derive the expression of the protocol:
\begin{equation}
\lambda_{i} = \dot{Q}_{i} + \sum_{j}M_{ij}Q_{j}
\label{eq:l}
\end{equation}
\begin{equation}
\dot{\lambda}_{i}=\ddot{Q}_{i} + \sum_{j}M_{ij}\dot{Q}_{j}
\label{eq:ld}
\end{equation}
We can thus write 
the Lagrangian as 
$$
\mathcal{L}\approx \mbox{const. }+ \mathcal{L}_{1} + \mathcal{L}_{2} + \mathcal{L}_{3} + \mathcal{L}_{4}
$$
where
$$
\mathcal{L}_{1} =\sum_{i=1}^{N} \nu^{i}Q_{i}\left(\dot{Q}_{i} + \sum_{j}M_{ij}Q_{j}\right)
$$
$$
\mathcal{L}_{2} =\sum_{i=1}^{N} \alpha^{i}\dot{Q}_{i}\left(\dot{Q}_{i} + \sum_{j}M_{ij}Q_{j}\right)
$$
$$
\mathcal{L}_{3} =\sum_{i=1}^{N} \gamma^{i}Q_{i}\left(\ddot{Q}_{i} + \sum_{j}M_{ij}\dot{Q}_{j}\right)
$$
$$
\mathcal{L}_{4} =\sum_{i=1}^{N} \eta^{i}\dot{Q}_{i}\left(\ddot{Q}_{i} + \sum_{j}M_{ij}\dot{Q}_{j}\right)
$$
Next, we evaluate the integrals $\int_{0}^{\tau}dt\mathcal{L}_{k}$ for $k=1,2,3,4$.
$$
\int_{0}^{\tau}dt\mathcal{L}_{1} =\sum_{i}\left\{ \frac{\nu^{i}}{2}\Bigl[Q^{2}_{i}\Bigr]_{0}^{\tau} +\right.
$$
$$
+\left.M_{ii}\int_{0}^{\tau}dt(\nu^{i}Q_{i}^{2}) + \sum_{j\ne i}M_{ij}\int_{0}^{\tau}dt(\nu^{i}Q_{i}Q_{j})\right\}
$$
$$
\int_{0}^{\tau}dt\mathcal{L}_{2} =\sum_{i}\left\{  M_{ii}\int_{0}^{\tau}dt(\alpha^{i}\dot{Q}_{i}^{2})\right.
$$
$$
\left.
+\sum_{j\ne i}M_{ij}\int_{0}^{\tau}dt(\alpha^{i}\dot{Q}_{i}Q_{j})\right\}
$$
If $\mathbf{F}(\mathbf{q}) = -\boldsymbol{\nabla}V(\mathbf{q})$ for some scalar function $V(\mathbf{q})$, $M_{ij}$ (the stability matrix) is always symmetric and the second integrals in the last expression can be evaluated explicitly
$$
\sum_{j\ne i}M_{ij}\int_{0}^{\tau}dt(\alpha^{i}\dot{Q}_{i}Q_{j})=\alpha^{i}\left[\frac{1}{2}\sum_{j\ne i}M_{ij}Q_{i}Q_{j}\right]_{0}^{\tau}
$$
When $M_{ij}$ has no particular symmetry, we can always decompose it in a symmetric $M_{ij}^{S} = (M_{ij}+M_{ji})/2$ and anti--symmetric part, i.~e. $M^{A}_{ij}=(M_{ij} - M_{ji})/2$. The anti--symmetric part sums up to zero and the above integral becomes
$$
\left[\frac{\alpha^{i}}{4}\sum_{j\ne i}M_{ij}^{S}Q_{i}Q_{j}\right]_{0}^{\tau}
$$
In any case, both expressions are constants that only depend on the values of $Q_{i}$ at the boundary points. Thus, they play no role in the minimization of $\mathcal{W}$. Next,
$$
\int_{0}^{\tau}dt\mathcal{L}_{3} =\sum_{i}\left\{ \gamma^{i}\Bigl[Q_{i}\dot{Q}_{i}\Bigr]_{0}^{\tau} - M_{ii}\int_{0}^{\tau}dt\left(\frac{\gamma^{i}}{M_{ii}}\dot{Q}_{i}^{2}\right)+\cdots\right\}
$$
where ``$\cdots$'' denotes constant terms similar to the one discussed above. Finally,
$$
\int_{0}^{\tau}dt\mathcal{L}_{4}=\sum_{i}\left\{\frac{\eta^{i}}{2}\Bigl[\dot{Q}_{i}^{2}\Bigr]_{0}^{\tau}  +\right.
$$
$$
+\left.
M_{ii}\int_{0}^{\tau}dt\left(\frac{\eta^{i}}{M_{ii}}\dot{Q}_{i}^{2}\right)+\sum_{j\ne i}M_{ij}\int_{0}^{\tau}dt(\eta^{i}\dot{Q}_{i}\dot{Q}_{j})\right\}
$$
Putting all the results together and omitting all the constant terms that do not influence the minimization of the Lagrangian, we get Eq.~\eqref{eq:WFIN}. 

%************************************************************************************

%\bibliography
\bibliography{Biblio}
\end{document}